\documentstyle[12pt]{article}

\newcommand{\be}{\begin{equation}}
\newcommand{\ee}{\end{equation}}
\newcommand{\ba}{\begin{eqnarray}}
\newcommand{\ea}{\end{eqnarray}}

\newcommand{\tr}{\rm tr}

\begin{document}
\hoffset=-.4truein\voffset=-0.5truein
\setlength{\textheight}{8.5 in}
\begin{titlepage}
\begin{center}
\hfill {LPTENS 07-49 }\\
\vskip 0.6 in
{\large \bf Intersection numbers of Riemann surfaces  from Gaussian matrix models}
\vskip .6 in
\begin{center}
{\bf E. Br\'ezin$^{a)}$}{\it and} {\bf S. Hikami$^{b)}$}
\end{center}
\vskip 5mm
\begin{center}
{$^{a)}$ Laboratoire de Physique
Th\'eorique, Ecole Normale Sup\'erieure}\\ {24 rue Lhomond 75231, Paris
Cedex
05, France. e-mail: brezin@lpt.ens.fr{\footnote{\it
Unit\'e Mixte de Recherche 8549 du Centre National de la
Recherche Scientifique et de l'\'Ecole Normale Sup\'erieure.
} }}\\
{$^{b)}$ Department of Basic Sciences,
} {University of Tokyo,
Meguro-ku, Komaba, Tokyo 153, Japan. e-mail:hikami@dice.c.u-tokyo.ac.jp}\\
\end{center}     
\vskip 3mm         

{\bf Abstract}                  
\end{center}
 We consider a Gaussian random matrix theory in the presence of an external matrix source. This matrix model,  after duality (a simple version of the closed/open string duality), yields  a generalized Kontsevich model through an appropriate tuning of the external source. 
The  n-point correlation functions of this theory are shown to provide  the 
intersection numbers of the moduli space of curves  with a p-spin structure, n marked points and top Chern class. This sheds some light on
Witten's conjecture on the relationship with the pth-KdV equation.
 \end{titlepage}
\vskip 3mm

\section{Introduction}

   The intersection theory for the moduli space of  curves with marked points has been studied by
various methods with the aim of  proving  Witten's conjectures \cite{Witten1} on the relationship between the structure of this space with  KdV flows. In particular this has  led to a solution through   Kontsevich's Airy matrix model\cite{Kontsevich}. 
The generalizations of this theory have been also considered also by Witten, who conjectured that
the intersection numbers of the moduli space with top Chern class is described by the $p$th-KdV equation \cite{Witten2}.

This generalization is related to  the twisted N=2 supersymmetric conformal field theory, and particulary it has a
close relation to the level $k = p-2$ WZW model \cite{Witten3}. 

In previous articles \cite{BH1,BH2}, we have found that a random matrix theory with an external source
gives an alternative  method for obtaining the intersection theory of the moduli space  of Riemann surfaces with a $p$-spin structure.

We have discussed it through an N-k duality  in the expectation values of the product of $k$  characteristic polynomials of $N\times N$ matrices . After tuning appropriately  the external
source to a critical value, we obtain the intersection numbers of the moduli space from the Fourier transforms of the
correlation functions. We had studied in the past  random matrix theory with an external source  as a way to tune  higher edge singularities, including  gap closing cases \cite{BH3,BH4}. Through this duality, plus tuning of the source,  we show here that one obtains the generalized Kontsevich  models.

In a previous article, we have computed the intersection numbers with one marked point for
 the generalized Kontsevich model
with $p$-spin curves by the replica method applied to a Gaussian random matrix theory.
We have also  found, by an alternative method, that the Fourier transform of the one point correlation function
$U(s)$ provides those intersection numbers and we have   found an explicit  agreement of the two approaches in the case of  one-marked point  and arbitrary genus  when $p=3$ \cite{BH2}.
 This was checking  
 that a Gaussian random matrix model with an external source is a dual (mirror) model to the generalized
Kontsevich models of $p$-spin structures with one marked point. (We will also show below  that it is true for $p>3$.)

For $p$=2, which is  Kontsevich original Airy matrix model,  the intersection numbers 
have been computed explicitly,
for a few marked points,  from a Fourier transform of the correlation functions at the edge 
of the spectrum of the density of eigenvalues,  in a random matrix model by Okounkov
\cite{OK1,OK2} 
and  in  our previous article
\cite{BH1}. Here we extend the analysis to the case of $p\ne 2$, again  from the edge singularities of the
correlation functions in a  matrix theory.

The intersection numbers are  characterized by the genus $g$ and marked points $n$ with a "spin-structure"  labelled by a
non-negative integer $j$, ($j= 0,1,..., p-1$).
Higher Airy matrix models for curves with a p-spin structure are given by \cite{Kontsevich}
\be\label{highAiry}
Z= \frac{1}{Z_0} \int dB {\rm exp}[ \frac{1}{p+1}{\tr} (B^{p+1} -\Lambda^{p+1}) - {\tr}(B-\Lambda)\Lambda^p]
\ee
where
\be
Z_0= \int dB {\rm exp}[\sum_{j=0}^{p-1} {\tr} \frac{1}{2} \Lambda^j B \Lambda^{p-j-1} B]
\ee

The free energy, the logarithm of the partition function $Z$,  is the generating function
for the intersection numbers $< \prod \tau_{m,j}>$,
\be\label{intersection}
F = \sum_{d_{m,j}}<\prod_{m,j} \tau_{m,j}^{d_{m,j}}> \prod_{m,j}\frac{t_{m,j}^{d_{m,j}}}{d_{m,j}!}
\ee
where 
\be
t_{m,j} = (-p)^{\frac{j-p - m(p+2)}{2(p+1)}}\prod_{l=0}^{m-1} (l p + j + 1){\tr}\frac{1}{\Lambda^{m p + j + 1}}
\ee
According to Witten \cite{Witten2} the intersection numbers is given by the top Chern class
$c_T$ and by the first Chern class $c_1$ as
\be\label{def}
< \tau_{m_1,j_1} \cdots \tau_{m_n,j_n} > = \frac{1}{p^g}\int_{\bar M_{g,n}} c_T(j_1,\cdots,j_n) 
\prod_{k=1}^n c_1({\mathcal{L}}_k)^{m_k}
\ee
with the condition which relates, for given $p$, the indices to the genus $g$ of the surface,
\be\label{grelation}
(p + 1) ( 2 g - 2 + n) = \sum_{i=1}^n ( p m_i + j_i + 1 ).
\ee

For more precise definitions and recent studies of the intersection numbers, we refer to the
literatures \cite{Jarvis,Faber,Shadrin,Hashimoto,LiuXu}.

The higher Airy matrix model of (\ref{highAiry}) is known to corresponds to a $(p - 1 )$ matrix model, which is conjectured to
satisfy
$p$th KdV equation. The Virasoro equations for this case are given in \cite{Dijk}. They have been used for computing recursively these intersection numbers ; here these numbers are given directly through a generating function.

\section{One point correlation function}

  In a previous article \cite{BH1} we have used 
explicit   integral representations for the correlation functions  
\cite{BH5,BH6,BH7} for a Gaussian unitary ensemble (GUE) of random matrices $M$ 
in the presence of an external matrix source.  

The probabililty distribution for $N\times N$ Hermitian matrices is 
\be\label{PA}
P_A(M) = \frac{1}{Z_A} e^{-\frac{N}{2}{\tr} M^2 - N {\tr} M A}
\ee

We denote the eigenvalues of the external source matrix by $a_\alpha$.

The Fourier transform of the one point correlation function, 
the density of state $\rho (\lambda)$,  is
\be
U(s) = \int \frac{d\lambda}{2\pi} e^{i \lambda s} <\tr \delta(\lambda - M) >
\ee
where the average $<\cdots>$ is taken with respect to  the probability  distribution $P_A(M)$.
An exact integral representation is known \cite{BH5,BH6,BH7} for arbitrary  $a_{j}$ : 
\be
U(s) = \frac{1}{N s} e^{\frac{N}{2} s^2} \oint \frac{du}{2\pi i}
\prod_{j=1}^{N} (1- \frac{s}{a_j - u})e^{N su}
\ee
In the following we shall  consider special cases in which the degeneracies of the distinct eigenvalues are proportional to $N$.  In particular we consider  $(p-1)$ different eigenvalues
$a_\alpha$, ($\alpha=1,...,p -1$),  each $a_\alpha$ 
being $\frac{N}{p-1}$ times degenerate. Then above expression reduces then to
\be \label{degenerate}
U(s) = \frac{1}{Ns} e^{\frac{N}{2}s^2} \oint \frac{du}{2\pi i} \prod_{\alpha=1}^{p-1}
(1 - \frac{s}{a_\alpha - u})^{\frac{N}{p-1}} e^{Nsu}.
\ee

We now consider the large-N limit, in a regime in which  $s$ and $u$ are of order $N^{-(p+1)}$. The distinct eigenvalues 
$a_\alpha$ are taken to be all of  order  one. Let us expand the integrand 
\be
U(s) = \frac{1}{Ns} e^{\frac{N}{2}s^2} \oint \frac{du}{2\pi i} {\rm exp}[
-\frac{N}{p-1}\sum_{\alpha=1}^{p-1} \sum_{m=0}^\infty (\frac{(u+s)^m}{m a_\alpha^m} - 
\frac{u^m}{m a_\alpha^m}) + N s u]
\ee
In the large-N limit the summation over $m$ may be truncated at $m\leq p+1$ since  higher orders vanish in that limit. 
We now specify the $(p-1)$ distinct eigenvalues of the external source by  the $(p-1)$  conditions :
\ba\label{equation}
&&\sum_{\alpha}^{p-1} \frac{1}{a_\alpha^2} = p-1, \hskip 5mm 
\sum_{\alpha=1}^{p-1} \frac{1}{a_\alpha^m} = 0, \hskip 5mm (m=3,4,...,p)\nonumber\\
&&\sum_{\alpha=1}^{p-1}\frac{1}{a_\alpha^{p+1}}\ne 0.
\ea
With these conditions, in the  regime of interest, one has
\be
U(s) = \frac{1}{N s} {\rm exp} [ - \frac{N}{p-1} s 
(\sum_{\alpha=1}^{p-1} \frac{1}{a_\alpha}) ] \int \frac{du}{2\pi i} {\rm exp}C [  u^{p+1} - (u+s)^{p+1} ]
\ee
with
\be
C = \frac{N}{p^2 - 1} \sum_{\alpha=1}^{p-1} \frac{1}{a_{\alpha}^{p+1}}.
\ee
The integration over $u$ takes a more symmetric form after the shift $u\rightarrow u - \frac{1}{2} s$ : 
\be\label{Us}
 U(s) = \frac{1}{N s} e^{- \frac{N s}{p-1}\sum_{\alpha=1}^{p-1} \frac{1}{a_\alpha}}\int \frac{du}{2\pi i}
e^{- C f(u)}
\ee
with
\ba
f(u) &=& (u + \frac{1}{2}s)^{p+1} - (u-\frac{1}{2} s)^{p+1}\nonumber\\
&=& \sum_{m=0}^{p+1} \left(\begin{array}{cc} p +1\\ m \end{array}\right)
 \frac{s^m}{2^m}(1 - (-1)^m) u^{p+1-m}
\ea
where the interval of the integration of $u$ is $(-i \infty, + i \infty)$.

\section{ Edge singularities}

In the large N limit, the density of state $\rho(\lambda)$ has a finite support, and thus develops singularities at the edge of the distribution. We consider now
the nature of this singularity for the $p$-spin case.

In the large N-limit 
the Green function $G(z) = \frac{1}{N}\langle \rm{tr} \frac{1}{z-M}\rangle$ is given  by a simple equation due to Pastur \cite{Pastur}   (see also \cite{BHZ})
\be  G(z) = \frac{1}{N} \sum_{\alpha=1}^{N} \frac{1}{z - a_\alpha - G(z)}.\ee
For a source which consists of $(p-1)$ distinct eigenvalues, each of them degenerate $N/(p-1)$ times,  this reads 
\be
G(z) = \frac{1}{p-1} \sum_{\alpha=1}^{p-1} \frac{1}{z - a_\alpha - G(z)}
\ee
If we expand this resolvent in  powers of $\frac{1}{a_\alpha}$, and use the conditions (\ref{equation}), we find
\ba
G &=& - \frac{1}{p-1}(\sum_{\alpha=1}^{p-1} \frac{1}{a_\alpha}) - 
\frac{1}{p-1}(z-G(z))(\sum_{\alpha=1}^{p-1} \frac{1}{a_\alpha^2})
\nonumber\\
&-& \frac{1}{p-1}(\sum_{\alpha=1}^{p-1}\frac{1}{a_\alpha^{p+1}})(z-G)^p
\ea
in which we have neglected terms of order  $(z-G)^{p+1}$ or higher.
From the first  condition  (\ref{equation}), we have
\be
z' = - \frac{1}{p-1}( \sum_{\alpha=1}^{p-1} \frac{1}{a_\alpha^{p+1}}) (z' - G'(z'))
\ee
where $z' = z + \frac{1}{p-1}\sum \frac{1}{a_\alpha}$ and 
$G'(z') = G(z)+ \frac{1}{p-1}\sum \frac{1}{a_\alpha^{p+1}}$.
The Green function has  a singularity proportional to  $z'^{1/p}$ ; thus  the density of state $\rho(\lambda)= -\frac{1}{\pi}
{\rm Im} G(\lambda)$ has an edge singularity characterized by an exponent $\frac{1}{p}$.

Note that  the conditions (\ref{equation}) for the $a_{\alpha}$ admit several solutions. 
However, all these different choices lead to the same singular behavior. In other words, the singularity exponent
is independent of the location of the critical points $z_c= -\frac{1}{p-1}\sum \frac{1}{a_\alpha}$.

For instance, in the case $p=3$, the (complex)solutions are $$(a_1,a_2)= \pm (1,- 1)$$ and
$$(a_1,a_2)=
\pm (\frac{\sqrt{3}}{2\sqrt{2}}+\frac{1}{2\sqrt{2}} i,\frac{\sqrt{3}}{2\sqrt{2}}-\frac{1}{2\sqrt{2}} i)$$.

The dual higher Airy matrix model in (\ref{highAiry}) is obtained from expectation values of characteristic
polynomials \cite{BH1,BH2}, leading to
\be
<\prod_{\alpha=1}^{p-1} {\rm det}(a_\alpha - i B)^{\frac{N}{p-1}}>=< {\rm exp}[ \sum_{\alpha=1}^{p-1}
{\tr} {\rm log}(1 - \frac{i B}{a_\alpha}) + N \sum {\rm log}(\prod_{\alpha=1}^{p-1} a_\alpha)]>
\ee
Expanding the logarithm,  with the conditions (\ref{equation}), we obtain
in the large N limit, the higher Airy matrix model (\ref{highAiry}) as explained in \cite{BH1,BH2}
(up to a change of the normalization factor).

\section{Intersection numbers of one marked point for $p$-spin curves}

We now derive the intersection numbers of one marked point from the asymptotic series
expansion of $U(s)$. 

{\bf (i) p=2}
We begin with the simple edge of the semi-circle law. In the large N-limit, in the range in which $s$ is of order $N^{-1/3}$, we have
\ba
U(s) 
&=& \frac{1}{Ns} e^{- \frac{C}{4} s^3} \int_{-\infty}^\infty \frac{du}{2\pi i} e^{3 C s u^2}\nonumber\\
&=& \frac{1}{Ns}\sqrt{\frac{\pi}{- 3Cs}}e^{-\frac{C}{4}s^3}
\ea
where $C= - \frac{N}{3}$.
By the change of the normalization due to the higher Airy matrix model of (\ref{highAiry}), we have
${\tilde s}^3/24 =  - N s^3/12$, and we get
\be
U = \frac{1}{Ns} \sqrt{\frac{\pi}{- 3C s}}\sum_{g=0}^\infty \frac{{\tilde s}^g}{(24)^g g!}
\ee
Noting that $s$ is a conjugate variable to $\Lambda$, this yields a series expansion in inverse powers of
$\Lambda$. From the definition   (\ref{intersection}), we obtain the intersection
numbers for one marked point,
\be
<\tau_{3g+1}>_{g} = \frac{1}{(24)^g g!}.
\ee
which agrees with our previous result \cite{BH2} based on the replica method.

{\bf (ii) p=3}
The critical point now corresponds to a density of states whose support consists of two disconnected segments, in the limit in which the gap closes. 
The intersection numbers with one marked point for p=3 have been obtained,  for  arbitrary genus $g$,  in 
our previous replica article \cite{BH2} :
\be\label{tau3}
<\tau_{{\frac{8g-5-j}{3}},j}>_g = \frac{1}{(12)^g g! }\frac{\Gamma(\frac{g+1}{3})}{\Gamma(\frac{2-j}{3})},
\ee
where the spin-index is $j=0$ for $g=3m+1$ and $j=1$ for $g=3m$ (m=1,2,3,...).

Near the edge point, $z_c= - \frac{1}{p-1}\sum \frac{1}{a_\alpha}$, we find in the scaling region from (\ref{Us})  
\be
U(s) = \frac{1}{Ns (3Ns)^{1/3}}Ai(\zeta)
\ee
where $\zeta= - N^{2/3}(4\cdot 3^{1/3})^{-1}s^{8/3}$
which was  already  established in  \cite{BH2}.
From the   standard asymptotic expansion of  Airy functions $Ai(\zeta)$, we have  two distinct series,
\ba\label{Aiseries}
U(s) &=& \frac{1}{N s (3 N s)^{1/3}}[ Ai(0)(1 + \frac{1}{3!}\zeta^3 + \frac{1\cdot 4}{6!}\zeta^6 + \frac{1\cdot 4\cdot 7}{9!}\zeta^9 +
\cdots)\nonumber\\
&& + Ai'(0)( \zeta + \frac{2}{4!} \zeta^4 + \frac{2\cdot 5}{7!} \zeta^7 + \frac{2\cdot 5\cdot 8}{10!}\zeta^{10}+ \cdots)]
\ea
where $Ai(0) = 3^{-2/3}/\Gamma(\frac{2}{3})$ and $Ai'(0) = - 3^{-1/3}/\Gamma(\frac{2}{3})$.

The first series in (\ref{Aiseries}) 
gives the intersection numbers for $j=1$. From this  series , noting that $s\sim \frac{1}{\Lambda^3}$, we
have
\ba
&&<\tau_{6,1}>_{g=3} = \frac{1}{(12)^3 3!}\cdot \frac{1}{3}, \hskip 5mm
<\tau_{14,1}>_{g=6} = \frac{1}{(12)^6 6!}\cdot \frac{4}{9},\nonumber\\
&&<\tau_{22,1}>_{g=9} = \frac{1}{(12)^9 9!}\cdot \frac{1\cdot 4\cdot 7}{3^3}, ...
\ea
From the second series, we have
\ba
&&<\tau_{1,0}>_{g=1} = \frac{1}{12}, \hskip 5mm
<\tau_{9,0}>_{g=4} = \frac{1}{(12)^4 4!}\cdot \frac{2}{3},\nonumber\\
&&<\tau_{17,0}>_{g=7} = \frac{1}{(12)^7 7!}\cdot\frac{2\cdot 5}{3^2},
\hskip 3mm
<\tau_{25,0}>_{g=10} = \frac{1}{(12)^{10}10!}\cdot\frac{2\cdot 5\cdot 8}{3^3}.
\ea
These values agree with (\ref{tau3}) and with the results obtained by completely different
mothods \cite{Shadrin}.

{\bf (iii) p = 4}

In this case, the critical values of the $a_\alpha$, which satisfy the conditions (\ref{equation}),
are obtained as the zeros of a cubic equation,
\ba\label{cubicequation}
   &&b_\alpha= \frac{1}{a_\alpha}\hskip 3mm (\alpha=1,2,3)\nonumber\\
   &&(x - b_1)(x - b_2)(x - b_3) = x^3 + \beta x^2 + \gamma x + \delta = 0
\ea
with
\be
\beta^2 = 9 \pm 3\sqrt{6},\hskip 3mm
\gamma = \frac{1}{2}(\beta^2 - 3),\hskip 3mm
\delta = \frac{3}{2 \beta} (\beta^2-1)
\ee
There is one real solution, and two complex conjugate solutions for the  $a_\alpha$.
(Although an analytic expression for the $a_\alpha$ ,  solutions of the cubic equation 
(\ref{cubicequation}) exists , we give  here the numerical values). There are two classes: 
\ba
(a_1,a_2,a_3) &=& (\pm 0.52523,\pm 0.41127 \pm 0.46403 i, \pm 0.41127 \mp 0.46403 i), \nonumber\\
(a_1,a_2,a_3) &=& (\pm 1.0076, \mp 0.71801 \pm 0.33908 i, \mp 0.71801 \mp 0.33908 i)
\ea
In both cases, the density of state has one critical edge, at which it behaves as $\rho(\lambda) \sim
\lambda^{1/4}$. Note that , contrary to the critical gap closing $p$=3 case,  
 for $p$=4,
there is just one single edge, similar to the $p$=2 case.

We have
\ba
U(s) &=& \frac{1}{N s} e^{-\frac{N}{3}s(\sum \frac{1}{a_\alpha})- \frac{N}{4^2\cdot 15}s^5
C}\int_{-\infty}^\infty
\frac{dv}{2\pi}{\rm exp}[ - C(\frac{s}{3}v^4 - \frac{s^3}{6}v^2)]\nonumber\\
&=& \frac{1}{N s} e^{-\frac{N}{3}s(\sum \frac{1}{a_\alpha})- \frac{N}{4^2\cdot 15}s^5
C}\int_{0}^\infty
\frac{dy}{2\pi} y^{-\frac{3}{4}} e^{- y + s^{5/2}\sqrt{\frac{N C y}{12}}}
\ea
where $C= N \sum_{\alpha=1}^3 \frac{1}{a_\alpha^5}$.
In a series expansion in powers of $s$, we obtain
\ba
U(s) &=& (\frac{3}{4Ns})(\frac{NCs}{3})^{-\frac{1}{4}}e^{-\frac{Ns}{3}\sum\frac{1}{a_\alpha}}\cdot
e^{-\frac{NC}{4^2\cdot 15} s^5}\nonumber\\
&& \times [ \Gamma(\frac{1}{4})( 1+ \frac{1}{2!}\cdot \frac{s^5}{4}(\frac{NC}{12})
+ \frac{s^{10}}{4!} (\frac{NC}{12})^2 \frac{1\cdot 5}{4^2}+ \cdots)\nonumber\\
&& + s^{\frac{5}{2}}(\frac{NC}{12})^{\frac{1}{2}}\Gamma(\frac{3}{4})(1 + 
\frac{1}{3!}\frac{3 s^5}{4}(\frac{NC}{12})+ \frac{s^{10}}{5!} \frac{4\cdot 7}{4^2}(\frac{NC}{12})^2+\cdots)]
\nonumber\\
\ea
For this $p=4$ case, there is the  overall factor   $e^{-\frac{NC}{4^2\cdot 15} s^5}$.
It must also   be expanded  and combined with the series to find the relevant terms in the s-expansion.
We obtain then  the intersection numbers from this expression,  with  the scaling $\tilde s^5 = \frac{NC}{12} s^5$,
\ba
&&<\tau_{1,0}>_{g=1} = \frac{1}{8},\hskip 3mm
<\tau_{3,2}>_{g=2} = \frac{9}{8^2\cdot 5!},\nonumber\\
&&<\tau_{6,0}>_{g=3} = \frac{9}{8^3\cdot 5!},\hskip 3mm
<\tau_{8,2}>_{g=4} = \frac{7\cdot 11}{8^5\cdot 5! \cdot 10},...
\ea
The results up to order $g=4$ had been computed in our replica article \cite{BH2}  
and indeed agree with
these new results.

{\bf (iv) $p \geq 5$}

In the case $p=5$ , the solutions of the conditions  (\ref{equation}) fall in  three different classes. (a) symmetric solution;
$(a_1,a_2,a_3,a_4) = (\sigma + \rho i,\sigma - \rho i,-\sigma + \rho i,-\sigma - \rho)$ with
$\sigma=\pm0.776887,\rho=\pm0.321797$.\\
(b)  $a_j$  given by $\pm$(0.624916,-1.01437,0.53363 +0.473515 i, 0.53363-0.473515 i).\\
(c) $a_j$  given by $\pm$(0.280577+0.5117 i,0.280577-0.5117 i,0.433665 + 0.158861 i,0.433665-0.158861 i).

These three cases all give a closing gap singularity  with same same exponent, $\rho(\lambda)\sim |\lambda|^{1/5}$.
We have
\be
U(s) = \frac{1}{Ns} e^{-\frac{N}{4}(\sum \frac{1}{a_\alpha}) }
\int \frac{du}{2\pi}
e^{-N(\sum \frac{1}{a_\alpha^6})[ \frac{1}{4} s u^5 + \frac{5}{24} s^3 u^3 + \frac{1}{64}s^5 u]}
\ee
The intersection numbers may be obtained from this expression in complete analogy  with the $p$=3 case. The integrand for $p=5$ presents five Stokes lines and the spin-label $j=0,\cdots,4$ characterizes  the various domains with different asymptotic expansions.  For $p>5$ the situation is similar to those described above. 
\section{Several marked points}
Up to now, we have only considered surfaces with one marked point. For higher intersection numbers we have to look at k-point correlation functions. 
The Fourier transform of the k-point  function $U(s_1,...,s_k)$ is also known in closed form ; it is given by  the integral representation
\cite{BH7}.
\ba 
&& U(s_1,\cdots, s_k) = \frac{1}{n} \langle \rm{tr} e^{s_1B} \cdots \rm{tr} e^{s_k B} \rangle \\ \nonumber
&& = (-1)^{k(k-1)/2} e^{\sum_1^k \frac{s_i^2}{2}}\oint \prod_1^k \frac{du_i}{2i\pi}
 e^{\sum_1^k (u_is_i)} \prod_{i=1}^k \prod_{m=1}^n (1+\frac{s_i}{u_i - a_m})\det \frac{1}{u_i+s_i-u_j}
\ea
For the two-point function  (k=2), with the same degenerate external source used in  (\ref{degenerate}), one has
\ba
U(s_1,s_2) &=& e^{\frac{N}{2}(s_1^2+s_2^2)}\oint \frac{du_1 du_2}{(2\pi i)^2} 
\prod_{\alpha=1}^{p-1} (1 - \frac{s_1}{a_\alpha -u_1})^{\frac{N}{p-1}}(1 - \frac{s_2}{a_\alpha - u_2})^{\frac{N}{p-1}}
\nonumber\\
&&\times 
\frac{e^{N(s_1 u_1+ s_2 u_2)}}{(u_1-u_2+s_1)(u_1-u_2-s_2)}
\ea
At the edge singularity, we expand the integrand in  powers of $\frac{1}{a_\alpha}$, with again the
critical constraints  (\ref{equation}),
\ba\label{twogeneralr}
&&\prod_{\alpha=1}^{p-1}(1 - \frac{s}{a_\alpha - u})^{\frac{N}{p-1}}e^{\frac{N}{2}s^2+ N s u}\nonumber\\
&&\sim {\rm exp}[ - \frac{N}{p-1}(\sum \frac{1}{a_\alpha}) s - \frac{N}{(p^2-1)}\sum_{\alpha=1}^{p-1}
(\frac{(u+s)^{p+1} - u^{p+1}}{a_\alpha^{p+1}})]
\ea
By the shift $u\rightarrow u-\frac{1}{2}s$, we obtain 
\ba\label{twogeneralr2}
U(s_1,s_2) &=& \int \frac{du_1du_2}{(2\pi i)^2}e^{- \frac{N}{p-1}(\sum \frac{1}{a_\alpha}) s_1 
- \frac{N}{(p^2-1)}(\sum_{\alpha=1}^{p-1}\frac{1}{a_\alpha^{p+1}})
((u_1+\frac{s_1}{2})^{p+1} - (u-\frac{s_1}{2})^{p+1})}\nonumber\\
&&\times e^{- \frac{N}{p-1}(\sum \frac{1}{a_\alpha}) s_2 
- \frac{N}{(p^2-1)}(\sum_{\alpha=1}^{p-1}\frac{1}{a_\alpha^{p+1}})
((u_2+\frac{s_2}{2})^{p+1} - (u_2-\frac{s_2}{2})^{p+1})}\nonumber\\
&&\times \frac{1}{(u_1-u_2 + \frac{s_1+s_2}{2})(u_1-u_2-\frac{s_1+s_2}{2})}.
\ea

Similarly for  the critical $k$-point correlation function,  for an arbitrary value of $p$, we have
\ba
&&U(s_1,...,s_k) =  \frac{1}{(2\pi i)^k}\int
\prod_{i=1}^k d u_i e^{-\frac{N C}{p^2-1} \sum_{i=1}^k [(u_i + \frac{s_i}{2})^{p+1} - (u_i - \frac{s_i}{2})^{p+1}]}
\nonumber\\
&&\times \frac{e^{-\frac{N}{p-1}(\sum\frac{1}{a_\alpha})\sum s_i}}{{\rm det} (u_i -u_j + \frac{1}{2}(s_i+s_j))}
\ea
with
$C= \sum_{\alpha=1}^{p-1} \frac{1}{a_\alpha^{p+1}}$. The leading connected part is obtained by  the longest cycles (of length $k$) in the expansion of the  determinant.

We focus now on the case $p$=3  (for $p$=2, this integral has  been already evaluated
and the intersection numbers for two marked points for arbitrary genus thereby obtained in \cite{BH1}, (see also 
\cite{OK2,LiuXu}).

For the  case $p$=3, we consider the critical values $(a_1,a_2)=(1,-1)$ for simplicity. 
The singularity is located at the origin (the gap closing point).
We replace the denominator  by the integral,
\be
\int_0^\infty dx e^{-[u_1-u_2 + \frac{1}{2}(s_1+s_2)] x} = \frac{1}{u_1-u_2 + \frac{1}{2}(
s_1+s_2)}
\ee
Then we have
\ba
U(s_1,s_2) &=& - \frac{e^{\frac{N}{4}(s_1^4+s_2^4)}}{s_1+s_2}\int_0^\infty
dx (e^{\frac{1}{2}(s_1+s_2) x} - e^{-\frac{1}{2}(s_1+s_2) x})\nonumber\\
&&\times
\int_{-\infty}^\infty \frac{dv_1}{2\pi } e^{i N s_1 v_1^3 + i(x- \frac{N}{4} s_1^3)v_1}
\int_{-\infty}^\infty \frac{dv_2}{2\pi }e^{iN s_2 v_2^3-i(x + \frac{N}{4}s_2^3) v_2}
\ea
which leads to
\ba
U(s_1,s_2) &=& \frac{1}{(\tilde s_1+ \tilde s_2)(3\tilde s_1)^{1/3}(3 \tilde s_2)^{1/3}}
\int_0^\infty dy {\sinh}(\frac{1}{2}(\tilde s_1 + \tilde s_2) y)
\nonumber\\
&&\times
Ai(\frac{y-\frac{1}{4}\tilde s_1^3}{(3\tilde s_1)^{1/3}})
Ai(-\frac{y+ \frac{1}{4} \tilde s_2^3}{(3\tilde s_2)^{1/3}})
\ea
where we have used the  scaled variables  $s_i = N^{-1/4} \tilde s_i$ and $x = N^{1/4} y$.
From now on we shall drop the tilde,  but all $s$'s should read $\tilde{s}$ instead. 

Note that this expression is, from its definition,  symmetric under the exchange of $s_1$ and $s_2$ although at this stage 
it looks asymmetrical.

If we scale $y$ as $y\rightarrow (3 \tilde s_1)^{1/3} y$, we obtain a 
 convenient  form for expanding in powers of $s_i$,
\ba\label{Taylor}
U(s_1,s_2) &=& \frac{1}{(s_1+s_2)(3 s_2)^{1/3}} \int_0^\infty dy
{\rm sinh}( \frac{1}{2}(s_1+s_2) (3 s_1)^{1/3} y)\nonumber\\
&& Ai( y - \frac{1}{4\cdot3^{1/3}}s_1^{8/3}) Ai( - (\frac{s_1}{s_2})^{1/3} y - \frac{1}{4\cdot
3^{1/3}}s_2^{8/3})
\ea
Note that $s_i$ is conjugate to $\Lambda_i$ in a Fourier transform. However, we have defined the dual external source
model with a term $\rm{Tr} (B \Lambda)$. The definition (\ref{highAiry}) of  the higher Kontsevich-Airy  models involves a different  power of $\Lambda$, a power 3
for  $p=3$.
Thus the  scaling is here given by 
\be\label{slambda}
s_i \sim \frac{1}{\Lambda_i^3}
\ee
Expanding then for large $\Lambda_i$, i.e. small $s_i$, the leading term is obtained by replacing the $\sinh{X}$ by its first  coefficient $X$ and dropping the $s_i^{8/3}$ in the Airy functions. Note that the corrections involve $s^{8/3}$ and thus are of relative order $1/\Lambda^8$.

The leading term, which is order $1/\Lambda$,
is
\ba
U(s_1,s_2) &=& \frac{1}{2}(\frac{s_1}{s_2})^{\frac{1}{3}} \int_0^\infty dy y Ai(y)Ai(-(\frac{s_1}{s_2})y)\nonumber\\
&=&\frac{1}{2}\frac{(s_1 s_2)^{\frac{1}{3}} (s_1^{\frac{1}{3}}+s_2^{\frac{1}{3}})}{(s_1+s_2)}
 (- Ai'(0)Ai(0))
\ea
where $Ai(0)= 3^{-\frac{2}{3}}\frac{1}{\Gamma(\frac{2}{3})}$ and $Ai'(0)= -3^{-\frac{1}{3}}\frac{1}{\Gamma(\frac{1}{3})}$, and thus $ - Ai'(0)Ai(0) = \frac{\sqrt{3}}{6\pi}$. 
The calculation involves the differential equation $Ai''(y) = yAi(y)$ followed by integrations by parts.

The next order, which provides the intersection numbers for two marked points at genus one, involves three terms. The first one is the cubic correction $X^3/6$ to the linear term of  $\sinh{X}$ ;  the other two involve the Taylor expansion of the $s^{8/3}$  terms in the Airy functions. 
\ba
\Delta U^{(1)}(s_1,s_2) &=& \frac{1}{16} \frac{s_1 (s_1+s_2)^2}{(3 s_2)^{\frac{1}{3}}}\int_0^\infty dy
y^3 Ai(y)Ai(-(\frac{s_1}{s_2})^{\frac{1}{3}}y)\nonumber\\
&=&\frac{1}{4\cdot 3^{1/3}}(s_1 s_2)^{\frac{4}{3}} (Ai'(0))^2 + \frac{1}{8\cdot 3^{1/3}}s_1 s_2^{\frac{2}{3}}(s_2-s_1)
J
\ea
where $J$ is
\be
J= \int_0^\infty dy Ai(y)Ai(-(\frac{s_1}{s_2})^{\frac{1}{3}} y).
\ee

\ba
\Delta U^{(2)}(s_1,s_2) &=& -\frac{1}{8\cdot 3^{1/3}}
(\frac{s_1}{s_2})^{\frac{1}{3}} 
s_1^{\frac{8}{3}} \int_0^\infty dy
y Ai'(y) Ai(-(\frac{s_1}{s_2})^{\frac{1}{3}} y) \nonumber\\
&=& \frac{1}{8 \cdot 3^{1/3}}
(\frac{s_1}{s_2})^{\frac{2}{3}} \frac{s_1^{\frac{8}{3}} s_2}{s_1+s_2} (Ai'(0))^2
+ \frac{1}{8\cdot 3^{1/3}} \frac{s_1^3 s_2^{\frac{2}{3}}}{s_1+s_2} J
\ea

\ba
&&\Delta U^{(3)}(s_1,s_2)= -\frac{1}{8\cdot 3^{1/3}} (\frac{s_1}{s_2})^{\frac{2}{3}} s_2^{\frac{8}{3}}
(\frac{s_2}{s_1})^{\frac{1}{3}}\int_0^\infty dy y Ai(y)Ai'(-(\frac{s_1}{s_2})^{\frac{1}{3}} y)\nonumber\\
&&= \frac{1}{8\cdot 3^{1/3}}\frac{s_1^{\frac{2}{3}}s_2^3}{s_1+s_2}(\frac{s_2}{s_1})^{\frac{1}{3}}(Ai'(0))^2
- \frac{1}{8\cdot 3^{1/3}}(\frac{s_1}{s_2})^{\frac{1}{3}}\frac{s_1^{\frac{2}{3}} s_2^3}{s_1+s_2} J
\ea

The terms which are proportional to the integral $J$ cancel beautifully when one adds the three correction terms. The addition leads to 
\be
\Delta U(s_1,s_2) = \frac{1}{8\cdot 3^{1/3}} (s_1 s_2)^{\frac{1}{3}}(s_1^2+s_1 s_2 + s_2^2) (Ai'(0))^2
\ee
From the scaling to $\Lambda$, we have
\be
(s_1 s_2)^{\frac{1}{3}}(s_1^2+s_1 s_2 + s_2^2) \sim \frac{1}{\Lambda_1^7 \Lambda_2}+ \frac{1}{\Lambda_1^4 \Lambda_2^4}
+\frac{1}{\Lambda_1\Lambda_2^7}.
\ee
They correspond to  $t_{0,0}\sim \frac{1}{\Lambda}$, $t_{2,0} \sim \frac{1}{\Lambda^{7}}$ and $t_{1,0}\sim \frac{1}{\Lambda^4}$, 
and they give the intersection
numbers for two marked points.
\be
<\tau_{0,0}\tau_{2,0}>_{g=1} = <\tau_{1,0}^2> = \frac{1}{12}.
\ee
which agrees with the solution of the Virasoro equation for $p=3$ \cite{Dijk,Hashimoto}. Note that here instead of the  recursive calculation of these numbers used by previous authors, we have, with the integral representation (\ref{Taylor}),  a generating function for arbitrary genera of intersection numbers with two marked points.

\section{ The $p\to  - 1$ limit}

Up to now $p$, which characterizes the spin-structure, was an integer larger than one.
It is interesting to consider how the theory is modified when it is continued to negative values of $p$ 
and in particular when $p=-1$. Since the generalized Kontsevich model involves  $\frac{1}{p+1}{\rm tr} B^{p+1}$, the limit $p\rightarrow -1$ gives a logarithmic potential ${\rm tr}{\rm log} B$.
This logarithmic potential corresponds to the Penner model \cite{HZ,Penner} 
and it is known to be related to the
Euler character $\chi$ of the Riemann surfaces.

As we have discussed, the intersection numbers do depend upon $p$ ;  for instance, in the case
of one marked point, we have \cite{BH2}
\be
<\tau_{1,0}>_{g=1} = \frac{p - 1}{24}.
\ee
The analytic continuation over $p$ to  negative values is thus possible and in this case, it gives simply $<\tau_{1,0}>_{g=1} = -\frac{1}{12}$ in agreement with the result of  \cite{HZ} : $\chi(\Gamma_1^1) = -\frac{ 1}{12}$.

One can indeed consider continuing to negative values many of  those formulae. For instance  in our previous work  concerning  the replica limit, $n\rightarrow 0$ for $n\times n$ matrix $B$,
we had found
\be
{\rm limit}_{n\rightarrow 0} \frac{1}{n} < {\rm tr} B^l > = \frac{\Gamma(4l+1)}{4^l \Gamma(2l+2)}
\ee
which is indeed  finite, equal to $-1/3$,  in the limit $l\rightarrow -1$.

The case of $p=-1$ is particulary interesting since Witten \cite{Witten3} had pointed out that in the limit  $k\to-3$
in the level-$k$ gauged WZW model, the intersection numbers, defined by (\ref{def}),
become the integral of the top Chern class alone,
\ba\label{Euler}
Z_g &=& (-1)^g \int_{\bar M_g} c_T\nonumber\\
&=& - \chi(M_{g,j})
\ea
where $\chi$ is the Euler character of the manifold. 

If we compare our normalizations to Witten's  correspondence with the level-k gauged WZW model, our definition of $p$ is related to $k$ by 
\be
p= k+2
\ee
and the limit $k\to -3$  indeed corresponds to the $p=-1$ limit.

\section{ Discussion }

In this article we have shown that the intersection numbers of Riemann 
surfaces, the moduli space of surfaces with n marked points, endowed with a p-spin  structure,  are 
obtained from a Gaussian random matrix theory with external source,  at an edge 
point where the asymptotic density of eigenvalues exhibits a 
singular behavior $\rho(\lambda) \sim \lambda^{\frac{1}{p}}$. 
The Fourier transforms $U(s_1,...,s_k)$ of the k-point correlation functions
provide the intersection numbers through the conjugacy relation
$s_i \sim \frac{1}{\Lambda_i^{p}}$.

We have found an integral representation for $U(s_1,...,s_k)$ at the critical
point. Witten's conjecture is that the intersection numbers may be obtained recursively through
Virasoro equations. Our formula is instead  a closed expression for arbitary $p$ and arbitrary genus ;  it
also gives the possibility  to continue in $p$  to negative values
(for instance $p=-1$).
 It is amusing to observe that
the Fourier transform $U(s_1,...,s_k)$ has an expansion as products of powers of $s_i$, which
corresponds precisely to the expansion in $t_{n,j}$ of the free energy in the generalized Kontsevich model.

Recently, this $p$-th generalized Kontsevich model has been discussed as an effective 
theory of open strings between Liouville D-branes \cite{SeibergShih,Gaiotto}. The duality, which we have
discussed with external source, corresponds to the relation between closed string (gravity) and open
string (gauge theory) with cosmological constants $\Lambda$. The random matrix model with external source
gives thus a theory for the case of  D0 branes.

A possible extension of this work  would deal with  the orthogonal-symplectic Gaussian matrix models
with external source, which are also  mutually dual  \cite{BH8,BH9,BH10}. The time dependent case should also be
reconsidered \cite{BH7} at the light of  intersection numbers theory. This is all left to further work. 

\end{document}